\documentclass[preprint,preprintnumbers,nofootinbib,aps,superscriptaddress,12pt]{revtex4-1}
\usepackage{amsmath,amssymb,bm,epsfig}
\usepackage{color}
\usepackage{natbib}
\usepackage{hyperref} 
\usepackage{ulem}
\usepackage{graphicx}
\RequirePackage{lineno}

\newcommand{\tch}{T_{\textrm{ch}}}

\newcommand{\mub}{\mu_{B}}

\newcommand{\gs}{\gamma_{S}}

\newcommand{\sNN}{\sqrt{s_{\textrm{NN}}}}

\newcommand{\pip}{\pi^{+}}
\newcommand{\pim}{\pi^{-}}

\newcommand{\kap}{\rm{K}^{+}}
\newcommand{\kam}{\rm{K}^{-}}
\newcommand{\ks}{\rm{K}_{S}^{0}}

\newcommand{\pbar}{\overline{\rm p}}

\newcommand{\lbarb}{\bar{\Lambda}}
\newcommand{\lap}{\Lambda}
\newcommand{\xip}{\Xi^{-}}
\newcommand{\xibar}{\overline{\Xi}^{+}}
\newcommand{\obar}{\overline{\Omega}}

\newcommand{\dndy}{\textrm{d}N/\textrm{d}y}

\def \beq{\begin{equation}}
\def \eeq{\end{equation}}
\def \beqa{\begin{eqnarray}}
\def \eeqa{\end{eqnarray}}

\def \l{\left(}
\def \r{\right)}
\def \l{\left(}
\def \r{\right)}

\begin{document}

\title{{Freezeout conditions in proton-proton collisions\\ 
at the top RHIC and LHC energies}}

\author{Sabita Das}
\email{sabita@rcf.rhic.bnl.gov}
\affiliation{Institute of Particle Physics and Key Laboratory of Quark $\&$ Lepton Physics (MOE) , 
Central China Normal University, Wuhan-430079, China}
\affiliation{School of Physical Sciences,\\ National Institute 
of Science Education and Research, HBNI, Jatni-752050, India}

\author{Debadeepti Mishra}
\email{debadeepti.m@niser.ac.in}
\affiliation{School of Physical Sciences,\\ National Institute 
of Science Education and Research, HBNI, Jatni-752050, India}

\author{Sandeep Chatterjee}
\email{sandeepc@niser.ac.in}
\affiliation{School of Physical Sciences,\\ National Institute 
of Science Education and Research, HBNI, Jatni-752050, India}

\author{Bedangadas Mohanty}
\email{bedanga@niser.ac.in}
\affiliation{School of Physical Sciences,\\ National Institute 
of Science Education and Research, HBNI, Jatni-752050, India}

\begin{abstract}

The freezeout conditions in proton-proton collisions at $\sqrt{s_{\textrm{NN}}}=
200$, $900$ and $7000$ GeV  have been extracted by fits to the mean hadron 
yields at mid-rapidity within the framework of the statistical model of an ideal 
gas of hadrons and resonances in the grand canonical ensemble. The variation 
of the extracted freezeout thermal parameters and the goodness of the fits with 
$\sqrt{s_{\textrm{NN}}}$ are discussed. We find the extracted temperature and baryon 
chemical potential of the freezeout surface to be similar in p+p and heavy ion 
collisions. On the other hand, the thermal behaviour of the strange hadrons is 
qualitatively different in p+p as compared to A+A. We find an additional parameter 
accounting for non-equilibrium strangeness production is essential for describing 
the p+p data. This is in contrast to A+A where the non-equilibrium framework could 
be successfully replaced by a sequential and complete equilibrium model with an 
early freezeout of the strange hadrons.

\end{abstract}

\maketitle

\section{Introduction}\label{sec.intro}

The statistical model of non-interacting gas of hadrons and resonances at 
some volume $V$, temperature $T$ and conserved charge chemical potentials 
$\mu_B$, $\mu_Q$ and $\mu_S$ corresponding to the three conserved charges 
of QCD, namely baryon number $B$, electric charge $Q$ and strangeness $S$ 
have been remarkably successful in providing a good qualitative description 
of the mean hadron yields in heavy ion collision experiments across a wide 
range of beam energies from AGS to LHC~\cite{becatini, andronic, paprevsd}. 
This could possibly indicate a hadronic medium in thermal equilibrium prior 
to freezeout. However, the extracted thermal parameters indicate that the 
freezeout surface lies very close to the hadronization surface~\cite{hotQCDtrans,WBtrans}. 
This has led to the suggestion that the hadrons are directly born into 
equilibrium from the quark gluon plasma (QGP) phase instead of there being 
a microscopic collision mechanism for equilibration~\cite{stock}. A microscopic 
collision picture has been suggested by invoking contribution from Hagedorn 
resonances with exponential mass spectrum~\cite{noronha}. Recently, in yet 
another approach based on Unruh radiation, an universal freezeout temperature 
was suggested for systems starting from e+e, p+p to heavy ions~\cite{unruh}. 
Thus, in spite of the enormous phenomenological success of the thermal models, 
the microscopic understanding of such fast thermal equilibration is still an 
open question.

One crucial ingredient in the application of thermal models is the choice of the 
ensemble to treat the conserved charges. Ideally, conserved charges in an open 
system should be treated within a grand canonical ensemble (GCE) while those in a closed 
system should be treated canonically. Thus, $4\pi$ data should be treated canonically 
while for mid-rapidity measurements that represent an open system, grand canonical 
ensembles should be applicable. However, it is not so straightforward in the case of 
particle production in relativistic collisions. It is believed that even if the criteria 
for applicability of GCE, $VT^3>1$ holds true for the bulk 
of the produced particles, canonical suppression might still be required when the 
number of carriers of a specific conserved charge are few~\cite{clpp1}. For this reason, 
strangeness has been treated canonically in p+p collisions owing to the small system 
size for the $\sNN=200$ GeV mid-rapidity data at RHIC~\cite{clpp1}. 
 
It is interesting to test the framework of thermal models in small systems~\cite{smallsystem, 
smallsystem2}. Previously, thermal models have been used to describe particle yields in 
small systems with a fair degree of success~\cite{cfop2, smallsystem, clpp1, clpp2, begun}. 
It is a commonly accepted notion that in small systems the formation of a thermally 
equilibrated fireball through multiple scattering of its constituents is less likely than in 
A+A collisions. However, it has been demonstrated through explicit application of thermal models 
that the quality of description of the data is similar for different system sizes~\cite{becattinisystemsize}. 

In this paper, we will apply the thermal model on mid-rapidity data in p+p collisions at 
$\sNN=200$ (RHIC), 900 (LHC) and 7000 (LHC) GeV. We find the mid-rapidity data is 
described by the GCE at all the above $\sNN$. A comparative 
study of two different schemes of treating the strange hadrons in p+p collisions - either 
having a strangeness correlation volume different from the fireball volume or using a 
strangeness undersaturation factor $\gamma_S$ - yielded similar goodness of fits for both the 
schemes~\cite{smallsystem}. We have kept the strangeness conservation volume equal to the fireball 
volume, allowing non-equilibrium strangeness production only through the departure from unity of 
$\gamma_S$. At the LHC energies we have fixed the chemical potentials to zero.
 
\section{Thermal Model }\label{sec.mod}

In a single chemical freezeout scheme (1CFO) all the hadrons freezeout from the same surface characterised by 
a single $\l V, T, \mu_B, \mu_Q, \mu_S\r$. The particle multiplicities become:
\begin{equation}
N_{i} = \frac{g_{i}V}{2\pi^{2}}\sum_{k=1}^{\infty}(\pm1)^{k+1}\frac{m_{i}^{2}T}{k} 
 ~K_{2}\left( \frac{km_{i}}{T} \right)~\textrm{exp}~(\beta k \mu_{i} )\gamma_S^{k |S_i|},
\label{eqNVch4}
\end{equation}
where $V$ is the fireball volume, $g_{i}$ is the degeneracy, $m_{i}$ is the particle 
mass, and $K_{2}$ is second order Bessel function. $\beta=\frac{1}{T}$, where 
$T$ is the chemical freezeout temperature. The plus sign is for bosons and the minus 
sign is for fermions. The hadron chemical potential $\mu_i$ in case of complete 
chemical equilibrium can be written down in terms of $\mu_B$, $\mu_Q$ and $\mu_S$ as 
follows
\beq
\mu_i = B_i\mu_B + Q_i\mu_Q + S_i\mu_S\label{eq.mu}
\eeq
where $B_i$, $Q_i$ and $S_i$ are the baryon number, charge and strangeness of 
the $i$th hadron. It is a standard practice to extract $\mu_S$ and $\mu_Q$ from 
the following constraints
\beqa
\text{Net} S &=& 0\label{eq.nets}\\
\text{Net} B/\text{Net} Q &=& 1\label{eq.netq}
\eeqa
Eq.~\ref{eq.netq} is valid only for p+p collisions. In A+A collisions, the unity in 
the RHS of Eq.~\ref{eq.netq} should be replaced by $\sim2.5$.
The remaining parameters $\l V, T, \mu_B\r$ are extracted from fits to hadron yields. 
The total yield $N_{i}^{\textrm{tot.}}$ of the $i$th hadron include primordial yields 
(produced directly in the reaction) and secondary yields which are the feed-down from 
the decays of heavier resonances 
\begin{equation}
N_{i}^{\textrm{tot.}} = N_{i}^{\textrm{prim}} + \sum_{\textrm{states}~j}
N_{j}^{\textrm{prim}}~\textrm{B.R.}~(j\rightarrow~i) ,
\end{equation}
where $N_{i}^{\textrm{prim}}$ is primordial multiplicity of species $i$ and 
$\textrm{B.R.}~(j\rightarrow~i)$ is the  branching ratio of $j$ to $i$ through all possible 
channels. We have used the THERMUS code~\cite{thms} which is available publicly for 
the 1CFO analysis.

As seen in Eq.~\ref{eqNVch4}, there is one more parameter, $\gamma_S$ which is also treated 
as a free parameter and extracted from fits to data. It accounts for possible chemical 
nonequlibrium of strangeness in the fireball. In a complete equlibrium scenario, $\gamma_S=1$.

\section{Data Sets}\label{sec.data}

We have used the p+p collision mid-rapidity data sets at RHIC with $\sNN = 200 $ GeV~\cite{stpikp, ststrng, stphi} and at LHC 
with $\sNN = 900$ GeV~\cite{lhc9pikp, lhc9st} and 7 TeV~\cite{lhc7pikp, lhc7xiomg, lhc7phi}. 
The $\pi^{\pm}$ and $\Lambda$ are feed-down corrected from weak decays whereas (anti)protons 
at RHIC are inclusive. The data sets from LHC have $\pi^{\pm}$, $\textrm{p}$, 
$\pbar$ and $\Lambda$ that are feed-down corrected from weak decays. The details about 
the data sets used in this study are given in Table~\ref{tb.data}.

  \begin{table}[t]
  \begin{center}
  \begin{tabular}{|c|c|c|c|c|c|}
  \hline
  $\sNN$ (GeV) & Expt. & System & Particle yields~($\dndy$) & Antiparticle yields~($\dndy$) 
  & Ref.\\
  \hline
 200&STAR&p+p&$\pip: 1.44\pm0.11$&$\pim: 1.42\pm0.11$&\cite{stpikp}\\[-2mm] 
 &&&$\kap: 0.150\pm0.013$&$\kam: 0.145\pm0.013$&\cite{stpikp}\\[-2mm] 
 &&&p: 0.138$\pm$0.012&$\pbar: 0.113\pm0.010$&\cite{stpikp}\\[-2mm] 
 &&&$\lap: 0.0385\pm0.0036$&$\lbarb: 0.0351\pm0.0033$&\cite{ststrng}\\[-2mm] 
 &&&$\xip: 0.0026\pm0.0009$&$\xibar: 0.0029\pm0.001$ &\cite{ststrng}\\[-2mm] 
 &&&$\ks: 0.134\pm 0.011$&&\cite{ststrng}\\[-2mm] 
 &&&$\Omega+\obar: 0.00034\pm0.00019$&&\cite{ststrng}\\[-2mm] 
 &&&$\phi: 0.018\pm 0.003$&&\cite{stphi}\\[+1mm]
 \hline
 900&ALICE&p+p&$\pip: 1.493\pm0.0741$&$\pim: 1.485\pm0.0741$&\cite{lhc9pikp}\\[-2mm] 
 &&&$\kap: 0.183\pm0.0155$&$\kam: 0.182\pm0.0155$&\cite{lhc9pikp}\\[-2mm] 
 &&&p: 0.083$\pm$0.0063&$\pbar: 0.079\pm0.0063$&\cite{lhc9pikp}\\[-2mm] 
 &&&$\lap: 0.048\pm0.0041$&$\lbarb: 0.047\pm0.0054$&\cite{lhc9st}\\[-2mm] 
 &&&$\Xi^{-}+\overline{\Xi}^{+}: 0.0101\pm0.0022$&&\cite{lhc9st}\\[-2mm] 
 &&&$\ks: 0.184\pm 0.0063$&&\cite{lhc9st}\\[-2mm]  
 &&&$\phi: 0.021\pm 0.005$&&\cite{lhc9st}\\[+1mm]
 \hline
 7000&ALICE&p+p&$\pip: 2.26\pm0.1$&$\pim: 2.23\pm0.1$&\cite{lhc7pikp}\\[-2mm] 
 &&&$\kap: 0.286\pm0.016$&$\kam: 0.286\pm0.016$&\cite{lhc7pikp}\\[-2mm] 
 &&&p: 0.124$\pm$0.009&$\pbar: 0.123\pm0.01$&\cite{lhc7pikp}\\[-2mm] 
 &&&$\xip: 0.008\pm0.000608$&$\xibar: 0.0078\pm0.000608$ &\cite{lhc7xiomg}\\[-2mm] 
 &&&$\Omega: 0.00067\pm0.000085$&$\obar:
 0.00068\pm0.000085$&\cite{lhc7xiomg}\\[-2mm] 
 &&&$\phi: 0.032\pm 0.004$&&\cite{lhc7phi}\\
 \hline
  \end{tabular}
  \end{center}
  \caption{Details of the data sets used for fit with references.}
  \label{tb.data}
  \end{table}

\section{Results}\label{sec.res}

 \begin{figure}[!hbtb]
 \vspace{-0.35cm}
 \begin{center}
 \includegraphics[scale=0.72]{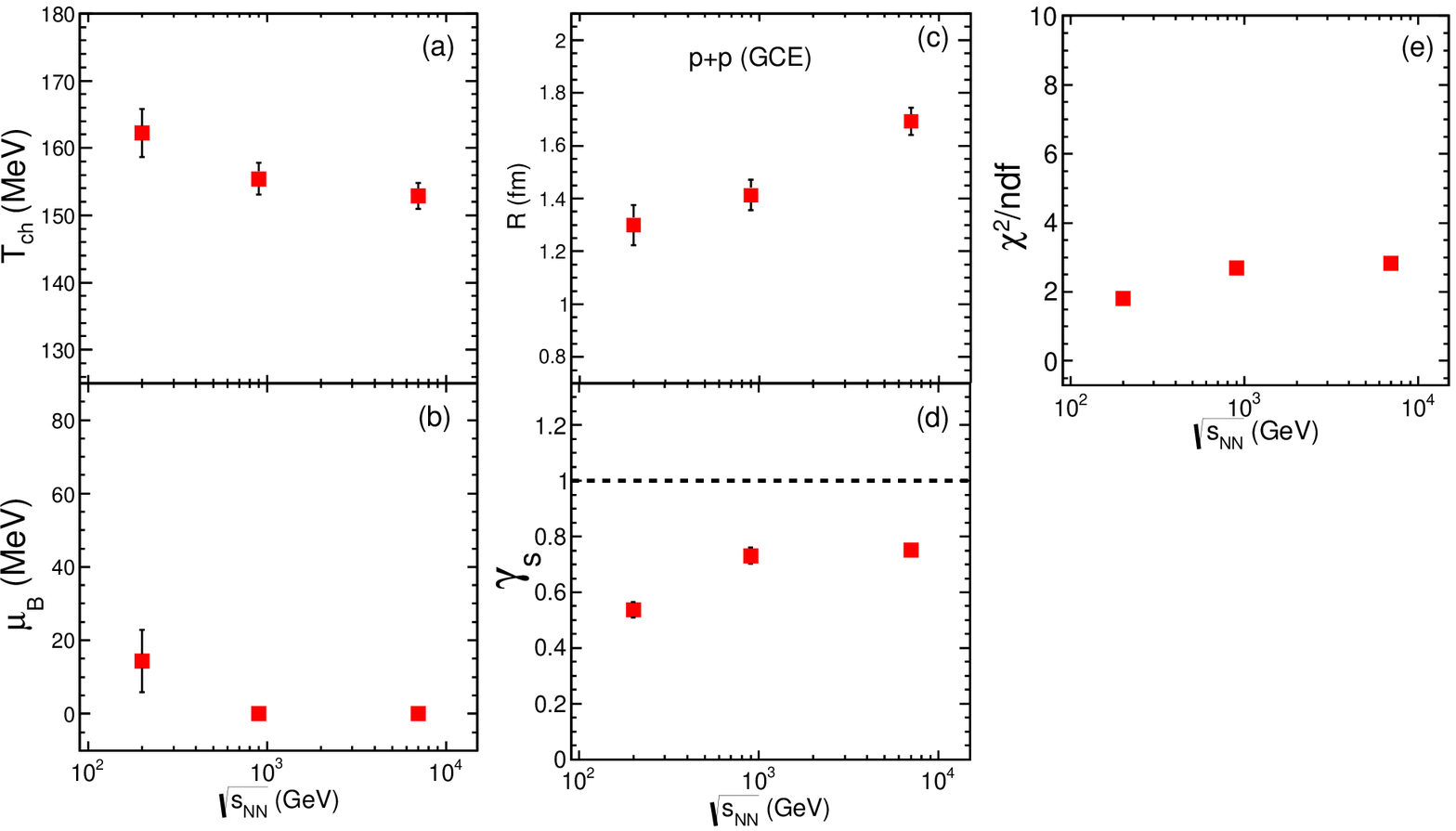}
 \vspace{0.2cm}
 \caption{(Color online) Freezeout parameters $\tch, \mub, \gs, R$ and $\chi^{2}/\textrm{ndf}$ 
 obtained from a statistical model fit~\cite{thms} using mid-rapidity particle yields.}
 \label{fig.1cfoensemble}
 \end{center}
 \vspace{-0.53cm}
 \end{figure}
 
In Fig.~\ref{fig.1cfoensemble} we have plotted the fitted freezeout parameters obtained in 
GCE in 1CFO. Previously, the RHIC $\sNN=200$ GeV p+p mid-rapidity data (excluding 
$\phi$) has been fitted in the strangeness canonical ensemble~\cite{clpp1}. However, we find here that the GCE 
provides reasonable description with $\chi^2/ndf\sim1-2$ even when including $\phi$.
The thermal model results for the mid-rapidity LHC data at $\sNN=900$ and 7000 GeV are new. The $\chi^2/ndf$ 
is around $2-3$ which has marginally increased from the top RHIC energy.

The value of $\gamma_S$ monotonically rises from $0.6$ to $0.8$ between RHIC and 
LHC energies. The freezeout $T$ on the other hand monotonically decreases from 160 MeV at $\sNN=$ 200 GeV to $\sim150$ 
MeV at 7 TeV. The fireball radius rises from $\sim1.3$ fm at $\sNN=200$ GeV to $\sim1.7$ fm at 
$\sNN=7$ TeV. $\mu_B$ is relatively flat and hovers around zero.

\begin{table}[t]    
\begin{center}
\begin{tabular}{ |c|c|c|c|c|c|c| } 
\hline
200 & 162.2 $\pm$ 3.6 & 14.4 $\pm$ 8.5 & 0.54 $\pm$ 0.03 & 1.3 $\pm$ 0.08 & 16.3 & 1.8 \\ 
\hline
900 & 155.4 $\pm$ 2.4 & 0.0 (Fixed) & 0.73 $\pm$ 0.03 & 1.42 $\pm$ 0.06 & 27.0 & 2.7 \\ 
\hline
7000 &152.9 $\pm$ 2.0 & 0.0 (Fixed) & 0.75 $\pm$ 0.02 & 1.69 $\pm$ 0.05 & 22.6 & 2.8 \\ 
\hline 
\end{tabular}
\end{center}
\caption{The chemical freezeout parameters extracted in 1CFO scheme in GCE at $\sNN=200, 
900$ and 7000 GeV.}
\label{tab.freezeoutparams}
\end{table}

\begin{figure}[!hbtb]
 \vspace{-0.35cm}
 \begin{center}
 \includegraphics[scale=0.40]{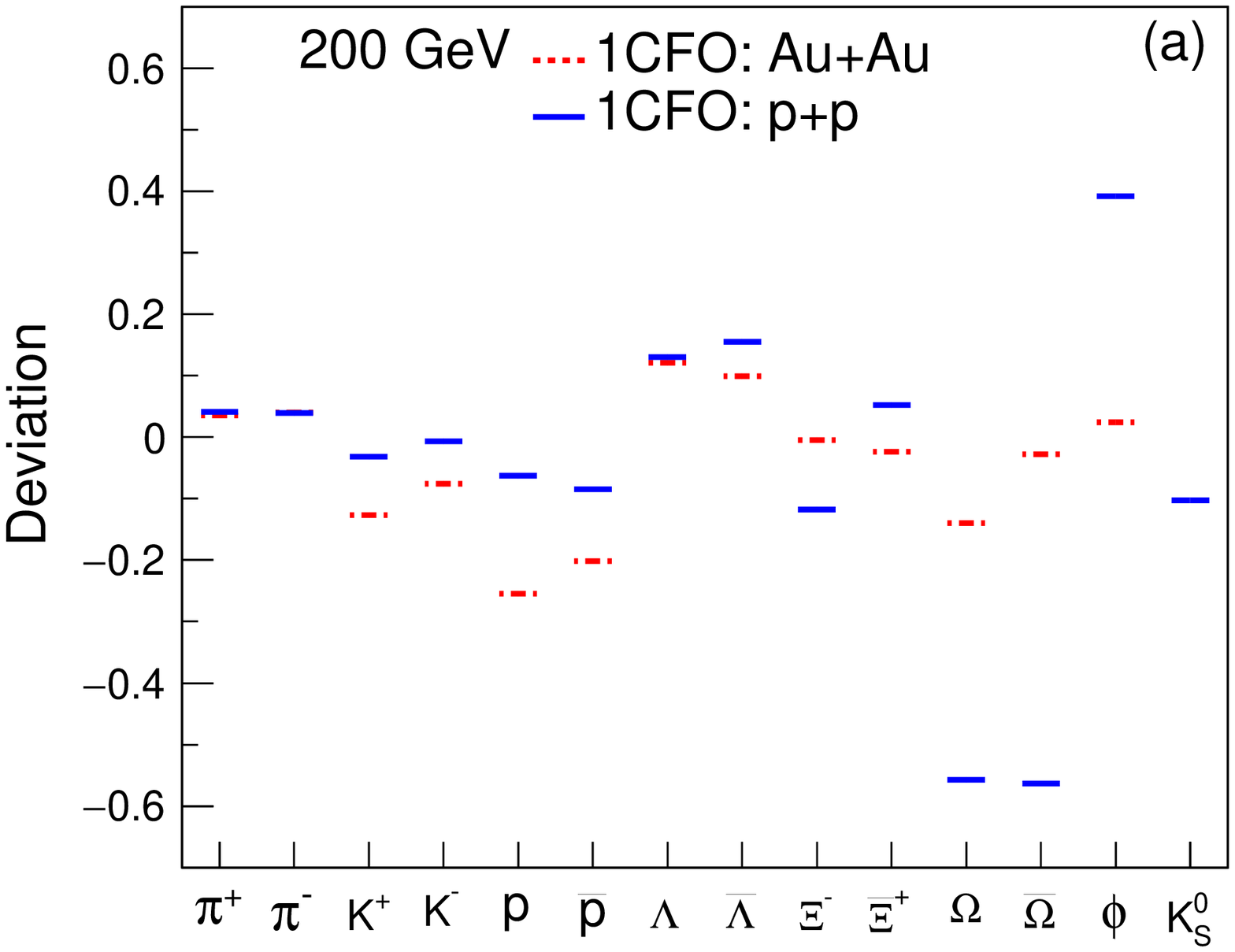}
 \includegraphics[scale=0.40]{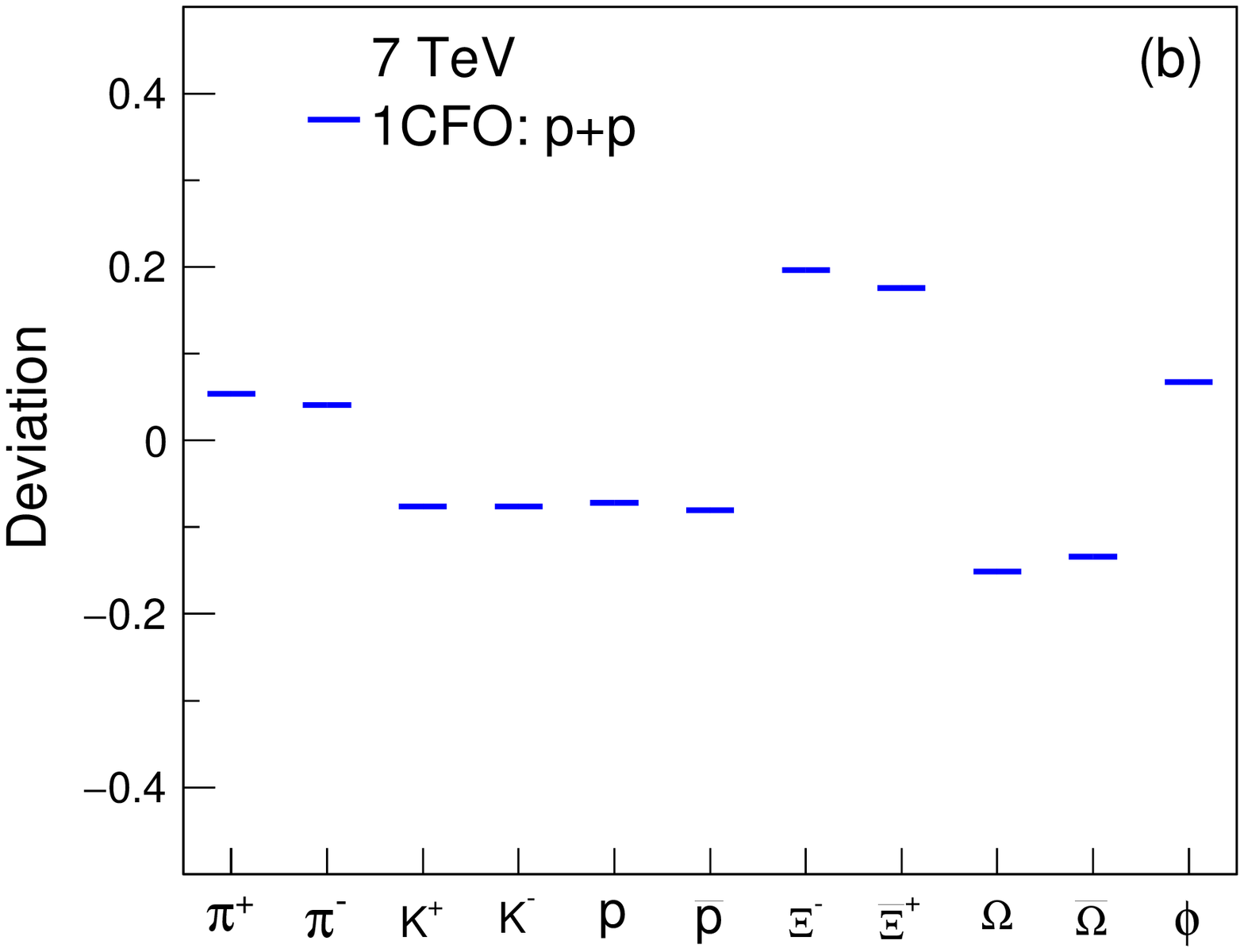} 
 \vspace{0.2cm}
 \caption{(Color online) Left: The deviation between model and data for each species in 1CFO at $\sNN=200$ GeV 
 compared between p+p and heavy ions. Right: The deviation between model and data in 1CFO at $\sNN=7$ TeV.}
 \label{fig.ppHIdev}
 \end{center}
 \vspace{-0.53cm}
\end{figure}

Earlier, we had noted that the $\chi^2/ndf$ marginally rises from the top RHIC to LHC energies. 
A rise in the $\chi^2/ndf$ do not necessarily mean a worsening of the thermal model fits. It 
could also occur due to more precise measurements. This could be verified by comparing the 
deviation between model and data defined as
\beq
\text{deviation} = \frac{\text{data} - \text{model}}{\text{data}}
\label{eq.dev}
\eeq
Figs.~\ref{fig.ppHIdev} (a) and ~\ref{fig.ppHIdev} (b) show the deviation for $\sNN=200$ 
and 7000 GeV respectively. 
At 7000 GeV, the deviation between data and model for all the hadron species lie within 
$20\%$. Even at 200 GeV, we find that except for $\Omega$ and $\phi$, the deviation for 
the rest of the hadrons are all within $20\%$. This shows clearly that the rise in $\chi^2/ndf$ 
from RHIC to LHC is due to more precision measurements at the LHC. Further, in 
Fig.~\ref{fig.ppHIdev} (a) we have also compared the deviation for each species between p+p 
and HICs for $\sNN=200$ GeV. We find that the hadrons with multiple valence strange quarks 
like $\phi$, $\Xi$ and $\Omega$ show higher deviation in p+p case compared to HIC.

\begin{figure}[!hbtb]
 \vspace{-0.35cm}
 \begin{center}
 \includegraphics[scale=0.35]{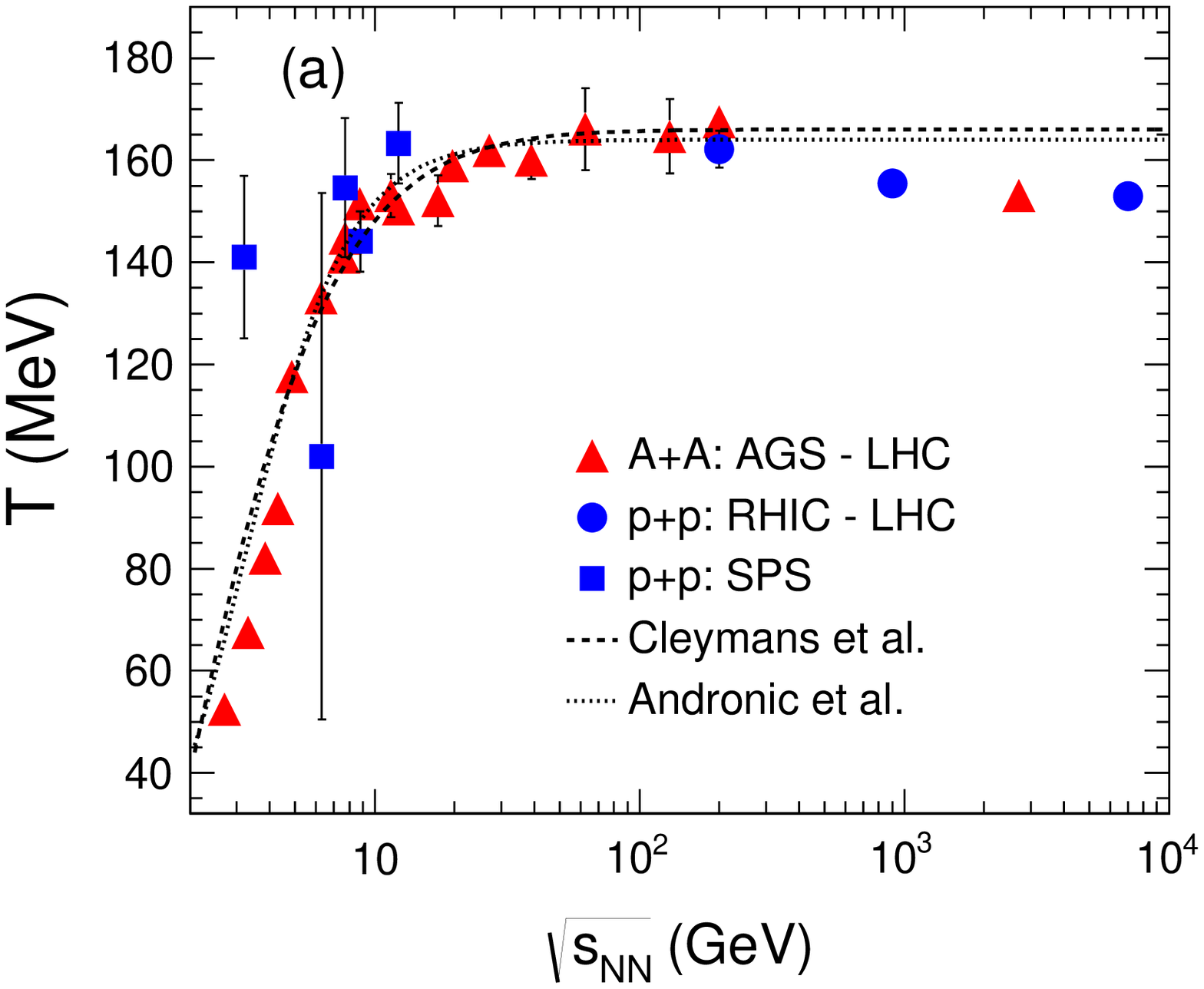}
 \includegraphics[scale=0.35]{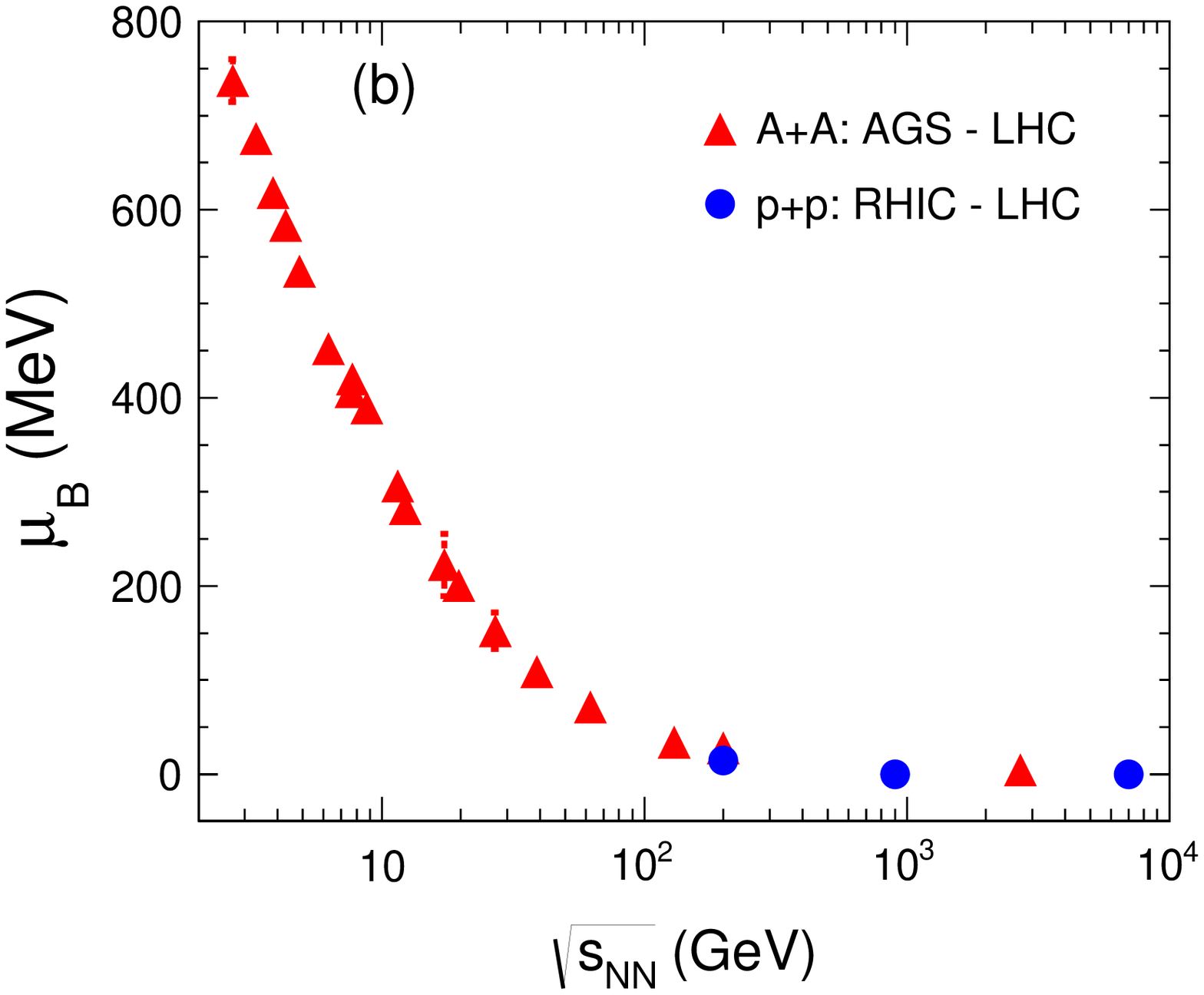}\\
 \includegraphics[scale=0.35]{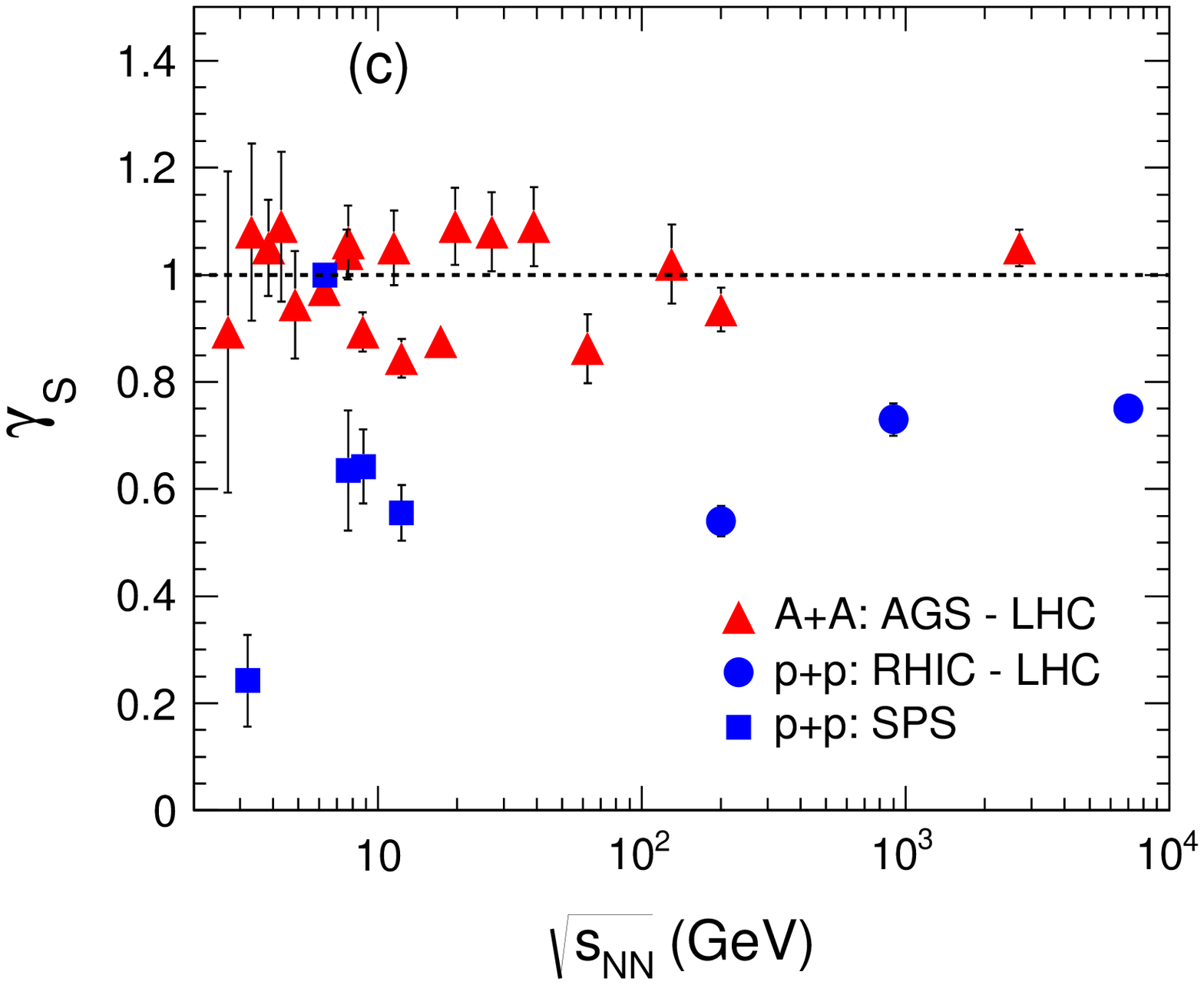}
 \includegraphics[scale=0.35]{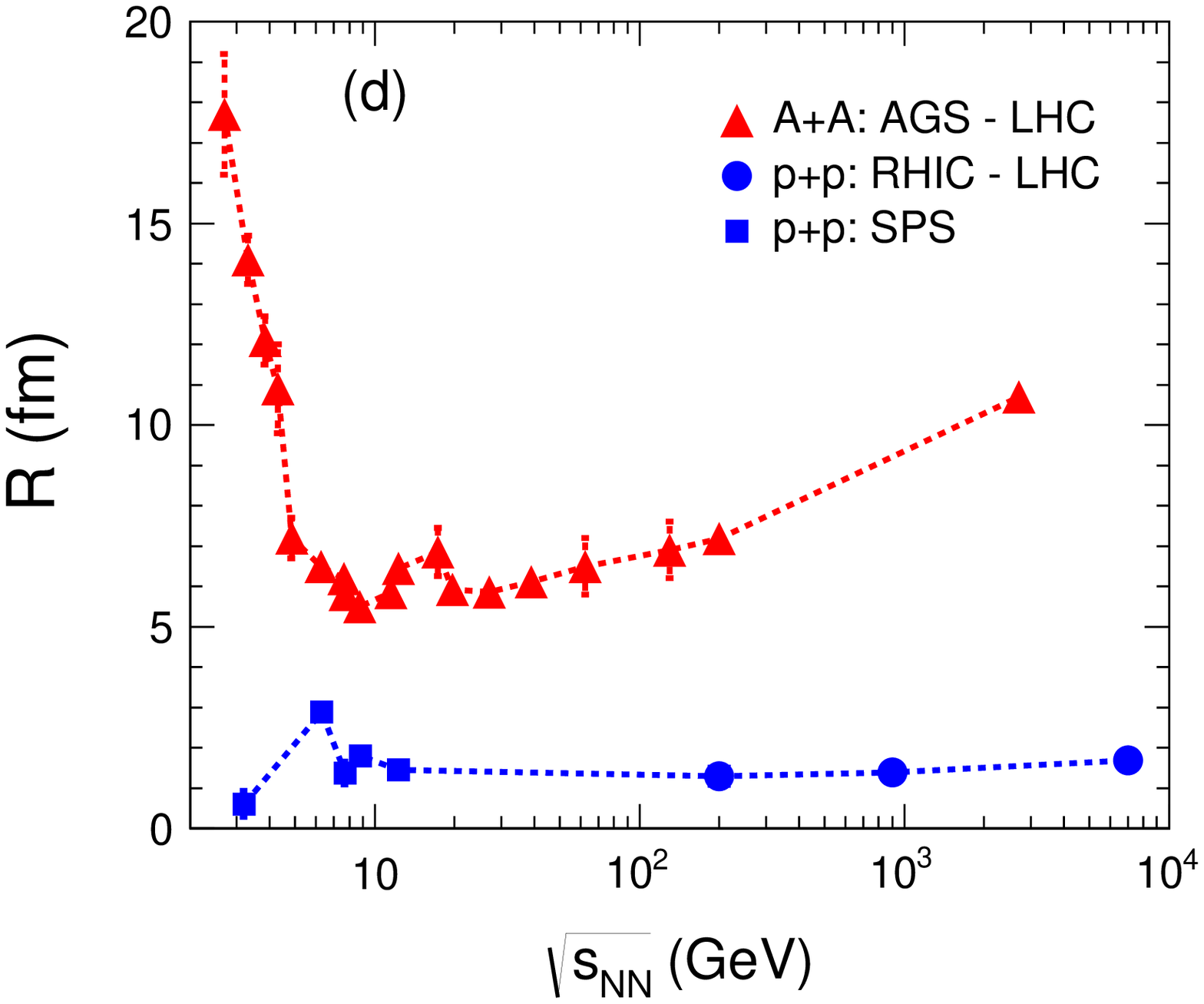}
 \vspace{0.2cm}
 \caption{(Color online) A compilation of $T$ (a), $\mu_B$ (b), $\gamma_S$ (c) and $R$ 
 (d) vs $\sNN$ in p+p collisions shown in blue squares at SPS energies taken from Ref.~\cite{begun} 
 and blue circles for RHIC and LHC energies which are the results of this paper. The results for A+A are shown 
 in red triangles for comparison~\cite{paprevsd}. The $T$ vs $\sNN$ parametrizations shown by dashed lines are from 
 Refs.~\cite{Tcleymans} and \cite{Tandronic}.}
 \label{fig.ppHITgammaS}
 \end{center}
 \vspace{-0.53cm}
\end{figure}

Finally, we have compared in Fig.~\ref{fig.ppHITgammaS} the freezeout parameters $T$, $\mu_B$, $\gamma_S$ 
and $R$ extracted in HICs with that of p+p in 1CFO. At lower $\sNN$, the p+p freezeout $T$ is higher than 
in A+A as was recently reported~\cite{begun}. However at higher beam energies ($\sNN>200$ GeV), the $T$ extracted 
in p+p is in agreement to that of HIC. As we go from RHIC to LHC energies, both p+p as well as A+A collisions show a 
decrease of the freezeout temperature by about 10 MeV. $\mu_B$ extracted from p+p is similar to that obtained 
from A+A. $\gamma_S$ and $R$ are quite different in the two systems. Between the top SPS and LHC energies, while 
the $R$ in A+A doubles, the corresponding rise in p+p is only about $20\%$. In this entire range, the 
radius in p+p is almost $5-10$ times smaller compared to A+A. In A+A collisions, $\gamma_S$ is consistent 
with unity while in p+p it is around 0.2 at SPS and then steadily rises before saturating around 0.8 at the LHC. 
This indicates significant strangeness suppression in p+p as compared to heavy ion even at the LHC energies. 
It will be interesting to see whether 
at even higher beam energies we produce strangeness in complete equilibrium or not. In this regard we note 
from Fig.~\ref{fig.ppHIdev} that in p+p there is large deviation between data and model as compared to heavy 
ion for hadrons with multiple strange valence quarks.

The recent data from Pb+Pb collisions at $\sNN=2.76$ TeV have renewed the 
interest in thermal models as the standard 1CFO freezeout scheme failed to explain 
the data satisfactorily, with a notable disparity between model and experiment in the proton to pion ratio, 
commonly known as the proton anomaly~\cite{panomaly,LHC}. Several alternative freezeout 
schemes have been proposed to address the above issue~\cite{alternatefreezeout1,noneq,
2cfothms,alternatefreezeout4}. One of them is the two freezeout scheme (2CFO) where 
those hadrons with non-zero strangeness content are 
allowed to freezeout at a different surface as compared to those with zero 
strangeness~\cite{2cfothms,alternatefreezeout4}. The 2CFO scheme has successfully 
described the proton anomaly~\cite{2cfothms} and transverse momentum 
spectra~\cite{2cfospectra} at LHC, the $\displaystyle{^3_{\Lambda}H\over ^3He}$ 
ratio at $\sNN=200$ GeV and $\bar{\Lambda}\over\bar{p}$ at lower beam energies 
which can not be described by the 1CFO scheme~\cite{paprevsd,2cfc2}. We have checked 
the above p+p data in the 2CFO scheme as well. However, unlike in HICs where the 2CFO scheme provides 
a much better description of the hadron yields than 1CFO~\cite{2cfothms}, here in p+p collisions we find the 
$\chi^2/ndf$ is similar and one does not gain much by introducing two additional parameters 
in 2CFO compared to $\gamma_S$ augmented 1CFO. Thus in p+p collisions, the 1CFO scheme with the additional strangeness 
suppresion factor $\gamma_S$ seems to be a better scheme than the complete chemical equlibrium but sequential 
freezeout scheme of 2CFO. The primary motivation for a 2CFO scheme in A+A collision 
is the expected flavor hierarchy in hadron-hadron cross-sections which result in different flavored hadrons 
freezing out at different times. However, in p+p collisions hadronic interactions are much reduced 
and the quick expansion results in a rapid freezeout leaving little room for sequential freezeout to 
occur.

\section{Summary}\label{sec.con}
The freezeout conditions for p+p collisions were extracted from the data on hadron yields at mid-rapidity 
for $\sNN=200$, 900 and 7000 GeV. Previous analyses have mostly focussed 
on a canonical treatment of strangeness in p+p collisions irrespective of the detector acceptance. We performed 
the analysis in the grand canonical ensemble as it is expected to describe the mid-rapidity system which 
behaves like an open system.

At these top beam energies, while the extracted temperature and baryon chemical potential is in agreement with those from heavy ion 
collisions, the strangeness suppresion factor comes out to be $\sim0.8$ in p+p. Thus the main difference arises in the freezeout 
condition for the strange hadrons. In A+A collisions, a complete thermal and chemical equilibrium scheme with 
early freezeout for strangeness provides a good description of the data. However, here in p+p collisions we 
found that a single freezeout scheme extended by a non-equilibrium factor for strangeness production provides the 
best description of the data amongst the different ensemble and freezeout schemes. We find a strong strangeness 
suppression across all the beam energies - about $20\%$ suppression is found even at the highest LHC energies. 
The expected shorter lifetime of the fireball in case of p+p collisions could be a reason behind such difference 
in the freezeout behaviour of the strange hadrons. 

\section{Acknowledgement}
We acknowledge Jurgen Schucraft for his helpful comments on the manuscript. SD thanks Jean Cleymans for helpful 
discussions and the MoST of China 973-Project No. 2015CB856901 and XIIth plan 
project no. 12-R$\&$D-NIS-5.11-0300 of Govt. of India for support. DM acknowledges the support from DAE and 
DST/SERB project of Govt. of India. SC thanks XIIth plan project no. 12-R$\&$D-NIS-5.11-0300 of Govt. of India 
for support. BM acknowledges support from DST SwarnaJayanti and DAE-SRC projects.

\end{document}